\ifdefined\HYDROSUPPLEMENT
  \documentclass[aps,prl,reprint,nofootinbib]{revtex4-2}
\else
  \documentclass[aps,prl,reprint,nofootinbib,superscriptaddress]{revtex4-2}
\fi

\usepackage{amsmath,amssymb,mathtools,mathrsfs}
\usepackage{graphicx}
\usepackage[dvipsnames]{xcolor}
\ifdefined\HYDROMAINONLY\usepackage{xr-hyper}\fi
\ifdefined\HYDROSUPPLEMENT\usepackage{xr-hyper}\fi
\usepackage{hyperref}
\usepackage[T1]{fontenc}
\usepackage[utf8]{inputenc}
\usepackage{CJKutf8}
\usepackage{tikz}
\usetikzlibrary{decorations.markings}
\AtBeginDocument{\setcounter{secnumdepth}{4}}

\newcommand{\ii}{\mathrm{i}}
\newcommand{\dd}{\mathrm{d}}
\newcommand{\cV}{\mathcal V}

\newcommand{\Adm}{\operatorname{Adm}}
\newcommand{\OP}{\operatorname{OP}}
\newcommand{\tp}[1]{\left[#1\right]_{+}}
\newcommand{\smeqref}[1]{%
  \ifdefined\HYDROMAINONLY\eqref{sm:#1}\else\eqref{#1}\fi}
\newcommand{\maineqref}[1]{%
  \ifdefined\HYDROSUPPLEMENT\eqref{main:#1}\else\eqref{#1}\fi}

\ifdefined\HYDROMAINONLY
\fi
\ifdefined\HYDROSUPPLEMENT
\fi

\begin{document}
\ifdefined\HYDROSUPPLEMENT
\title{Supplemental Material for ``The Hydrotope in the Water-Wave Action''}
\author{Qu Cao, Song He, Jirong Jing, and Qiupeng Li}
\noaffiliation
\maketitle
\else
\begin{CJK*}{UTF8}{gbsn}
\title{The Hydrotope in the Water-Wave Action}

\author{Qu Cao(曹趣)}
\email{caoqu@westlake.edu.cn}
		\affiliation{Department of Physics, School of Science, Westlake University, Hangzhou 310030, China}
\author{Song He(何颂)}
\email{songhe@itp.ac.cn}
\affiliation{New Cornerstone Laboratory, Institute of Theoretical Physics, Chinese Academy of Sciences, Beijing 100190, China}

\affiliation{School of Fundamental Physics and Mathematical Sciences,
Hangzhou Institute for Advanced Study, University of Chinese Academy of Sciences,
        Hangzhou 310024, China}
\author{Jirong Jing(景继荣)}
\email{jingjirong26@mails.ucas.ac.cn}

\affiliation{School of Fundamental Physics and Mathematical Sciences,
Hangzhou Institute for Advanced Study, University of Chinese Academy of Sciences,
Hangzhou 310024, China}

\author{Qiupeng Li(李秋鹏)}
\email{lieqiupeng@gmail.com}
\affiliation{School of Fundamental Physics and Mathematical Sciences,
Hangzhou Institute for Advanced Study, University of Chinese Academy of Sciences,
Hangzhou 310024, China}

\begin{abstract}
The Hydrotope gives a geometric representation of tree-level water-wave amplitudes with two negative spatial momenta as the volume of a hyperplane slice of a box. We trace the origin of this geometry directly to the water-wave action. Writing the
$n$-point contact interaction as
$\cV_n=\sum_{i<j}w_iw_jh_{ij}^{(n)}$, we show that when the two marked momenta have the same sign and every spectator has the opposite sign, the corresponding coefficient is $h_{ij}^{(n)}=2H_n$, where $H_n/(n{-}3)!$ is the Hydrotope volume. More generally, every fixed-pair coefficient admits a denominator-free ordered-flag representation as an oriented sum of $(n{-}3)$-dimensional box volumes. We then sum all two-minus trees by cutting
each at the unique vertex joining its two minus branches. The coefficients multiplying minus–minus, minus–plus, and plus–plus frequency bilinears reduce, respectively, to $(2^{n{-}1}{-}2)H_n$, $0$, and $2H_n$, and immediately reproduce the known amplitude $2^{n{-}1}w_1w_2H_n$. Thus, the Hydrotope—and a broader class of related box-slice geometries—is already encoded locally in the water-wave action. This points to further hidden simplicity and geometric structure in general water-wave amplitudes.
\end{abstract}
\maketitle
\end{CJK*}
    
\textit{Introduction.}---Striking geometric structures in scattering theory often become visible only after large classes of diagrams have been summed~\cite{Elvang:2013cua,Henn:2014yza,Travaglini:2022uwo,
Arkani-Hamed:2008owk,Arkani-Hamed:2012zlh,
Arkani-Hamed:2013jha,Arkani-Hamed:2017mur}.
More recently, machine-learning and AI-assisted methods have provided a complementary route to discovering such structures~\cite{Cheung:2024svk,Moynihan:2026mbz,
Guevara:2026qzd,Guevara:2026qwa}. A nonrelativistic setting in which similar phenomena arise is the theory of one-dimensional deep-water gravity waves. Its perturbative and diagrammatic formulations have a long history~\cite{Phillips,Hasselmann1966,Krasitskii}, and unexpected simplifications were already observed in five-wave interactions~\cite{DyachenkoLvovZakharov,Lvov}. More recently, tree amplitudes with two negative spatial momenta were shown to be proportional to the volume of a hyperplane slice of a box---the Hydrotope~\cite{Hydrotope}. In the conventional surface formulation, however, this geometry is far from manifest. Eliminating the fluid bulk produces a surface theory governed by a shape-dependent Dirichlet--Neumann operator (DNO)~\cite{CraigSulem,Lannes}. In the Lagrangian formulation, the inverse DNO appears explicitly, and its recursive expansion generates contact interactions of arbitrarily high multiplicity~\cite{Ussembayev}.

The Hydrotope was originally uncovered through AI-assisted analysis of complete tree amplitudes computed using Berends--Giele recursion~\cite{BerendsGiele,Hydrotope}. Here we show that the same geometry is already encoded locally in the water-wave action. We keep the frequencies independent, impose only spatial-momentum conservation, and decompose each contact vertex into coefficients multiplying frequency pairs. When the two marked momenta have the same sign and all spectator momenta have the opposite sign, the corresponding ordered action kernel collapses to a product of finite-difference operators acting on a single truncated power. This product is precisely the inclusion--exclusion formula for the volume of a hyperplane slice of a box. The ordinary Hydrotope is therefore already present in a single contact interaction, before the dispersion relation is imposed or any exchange diagrams are included. For arbitrary momentum-sign assignments, every fixed-pair coefficient admits a canonical, denominator-free ordered-flag representation as a finite oriented sum of box volumes, with the ordinary Hydrotope recovered in the pair-compatible chamber. The action thus encodes a broader geometric structure: every frequency-pair coefficient is a finite oriented sum of box volumes, even when it cannot be represented as the volume of a single convex body.

We then show how the complete two-minus tree sum selects and dresses this action-level geometry. Every generic tree contains a unique central vertex at which its two minus branches meet. Cutting the tree at this vertex replaces all attached subtrees by exact one-minus currents and converts the full diagrammatic sum into a finite sum over labelled partitions. Before frequency conservation is imposed, the coefficients multiplying minus--minus, minus--plus, and plus--plus frequency bilinears reduce, respectively, to
$(2^{n-1}{-}2)H_n$, $0$, and $2H_n$
Once the external legs are placed on shell, these terms combine into the known Hydrotope amplitude. The two-minus sector therefore provides an exact realization of a more general mechanism: the contact interaction supplies simple geometric building blocks, while tree dressing cancels the mixed-sign coefficients and reorganizes the surviving terms into a single Hydrotope function. This mechanism suggests that other momentum-sign sectors may admit comparably simple geometric organizations, even when their complete amplitudes are not themselves volumes of a single Hydrotope.

\textit{Geometry in a single contact interaction.}---We consider surface gravity waves, neglecting surface tension, on an inviscid,
incompressible, and irrotational fluid of infinite depth, with free surface
$y=\xi(x,t)$. The bulk velocity is $\nabla\phi$, with $\phi$ harmonic in the
fluid domain. Starting from Luke's variational principle and integrating out
the bulk degrees of freedom, the dynamics can be written entirely in terms of
the surface elevation $\xi$ and the boundary potential
$\psi=\left.\phi\right|_{y=\xi}$~\cite{Luke,Zakharov,CraigSulem,Lannes}. Let $G(\xi)$ denote the Dirichlet--Neumann operator, which maps $\psi$ to the normal derivative of $\phi$ at the free surface. The resulting surface
Lagrangian is
\begin{equation}
 \mathcal{L}
 =\langle\psi,\dot\xi\rangle
 -\frac12\langle\psi,G(\xi)\psi\rangle
 -\frac g2\langle\xi,\xi\rangle ,
 \label{eq:surface-action}
\end{equation}
where
$\langle f,g\rangle:=\int_{\mathbb R}\dd x\,f(x,t)g(x,t)$
denotes the spatial bilinear pairing, and we set $g=1$ henceforth. Eliminating $\psi$ using $\dot{\xi}=G(\xi)\psi$ gives the kinetic term
\begin{equation}
    \mathcal{L}_{\mathrm{kin}}
    =
    \frac{1}{2}
    \left\langle
        \dot{\xi},G(\xi)^{-1}\dot{\xi}
    \right\rangle .
    \label{eq:inverse-action}
\end{equation}

For deep-water surface gravity waves in one horizontal dimension
~\cite{CraigWorfolk,DyachenkoLvovZakharov,Lvov}, consider an $n$-field
contact interaction with momentum $k_i$ and frequency $w_i$ assigned to
each leg. We strip off the overall frequency-conserving delta function and
the common Feynman-rule phase, impose $\sum_i k_i=0$, and treat the
all-incoming signed frequencies as independent variables. Since each
interaction contains exactly two time derivatives, the momentum-space
contact vertex is quadratic in the frequencies and takes the form
\begin{equation}
 \cV_n=\sum_{1\leq i<j\leq n} w_i w_j\,
 h_{ij}^{(n)}(k_1,\ldots,k_n).
 \label{eq:contact-polarization}
\end{equation}

The inverse operator in Eq.~\eqref{eq:inverse-action} can be eliminated in one
spatial dimension using harmonic conjugation~\cite{DyachenkoConformal}. Let
$\psi$ denote the boundary value of a harmonic function in the fluid, and
$\widetilde\psi$ that of its harmonic conjugate. Applying the chain rule along
the moving surface gives
\begin{equation}
 \partial_x\widetilde\psi=-G(\xi)\psi,
 \qquad
 G(\xi)\widetilde\psi=\partial_x\psi.
 \label{eq:harmonic-pair}
\end{equation}
Eliminating $\widetilde\psi$ on nonzero-momentum modes then yields
\begin{equation}
 G(\xi)^{-1}=-\partial_x^{-1}G(\xi)\partial_x^{-1}.
 \label{eq:harmonic-duality}
\end{equation}
Defining $\chi=\partial_x^{-1}\dot\xi$, Eq.~\eqref{eq:inverse-action} becomes
\begin{equation}
 \mathcal{L}_{\rm kin}
 =\frac12\langle\chi,G(\xi)\chi\rangle .
 \label{eq:direct-action}
\end{equation}

At the flat surface, $G(0)=|\partial_x|$; with $g=1$, the quadratic part of
Eq.~\eqref{eq:direct-action} gives the deep-water dispersion relation
$w_i^2=|k_i|$. At order $n$, the two factors of $\chi$ select a frequency pair
$w_iw_j$, while the remaining $n{-}2$ fields arise from the expansion of
$G(\xi)$. We first evaluate this contact kernel in the sign sector relevant to
the two-minus amplitude. Let $k_1=-\alpha$, $k_2=-\beta$, and
$k_a=x_a>0$ for $a\in P=\{3,\ldots,n\}$, after possibly exchanging legs $1$ and $2$, with
$\alpha\geq\beta>0$. Momentum conservation then implies
$x_P=\alpha+\beta$. The Hydrotope is the slice
\begin{equation}
 \mathcal W_n:=\left\{(t_a)_{a\in P}:0\leq t_a\leq x_a,
 \quad \sum_{a\in P}t_a=\beta\right\}.
 \label{eq:hydrotope-slice}
\end{equation}

We normalize the slice volume as~\footnote{With this convention, $\operatorname{Vol}(\mathcal W_n)$ equals the intrinsic
Euclidean volume of the slice divided by $\sqrt{n-2}$.}
\begin{equation}
 \operatorname{Vol}(\mathcal W_n):=
 \int_{\prod_{a\in P}[0,x_a]}\!\dd^{n-2}t\,
 \delta\!\left(\beta-\sum_{a\in P}t_a\right).
 \label{eq:hydrotope-volume}
\end{equation}

 By the standard
box-spline inclusion--exclusion formula~\cite{BoxSplines},
\begin{equation}
 H_n:=(n-3)!\operatorname{Vol}(\mathcal W_n)
 =\sum_{A\subseteq P}(-1)^{|A|}
 \tp{\beta-x_A}^{n-3},
 \label{eq:hydrotope}
\end{equation}
where $x_A:=\sum_{a\in A}x_a$ and
$\tp{x}:=x\Theta(x)=\max(x,0)$.

The same formula can be derived directly from the action. Let
$\overline F(a,b;P)$ denote the spectator-symmetrized directed kernel obtained by recursively inverting the standard Craig--Sulem expansion of $G(\xi)$~\cite{CraigSulem,Lannes}. In one horizontal dimension,
Eq.~\eqref{eq:harmonic-duality} relates this kernel to that of
Eq.~\eqref{eq:direct-action} by harmonic conjugation~\cite{DyachenkoConformal}. Although the ordered inverse-DNO recursion
contains apparent subset-momentum denominators, each locally
momentum-conserving DNO subkernel supplies the corresponding endpoint
momenta and cancels them. After spectator symmetrization, the remaining
ordered insertions become commuting finite-difference operators. Hence, for
$0<z<x_P$, one finds
\begin{equation}
 \begin{aligned}
 \overline F\bigl(-(x_P-z),-z;P\bigr)
 &=-2\prod_{a\in P}\Delta_{x_a}\tp{z}^{n-3},\\
 (\Delta_xf)(z)&:=f(z)-f(z-x).
 \end{aligned}
 \label{eq:action-finite-difference}
\end{equation}

Let $\mathscr H_P(z):=\prod_{a\in P}\Delta_{x_a}\tp{z}^{n-3}$ denote the product of finite-difference operators, so
that $H_n=\mathscr H_P(\beta)$. The contact coefficient is obtained by
combining the two directed endpoint orderings,
\begin{equation}
 h_{12}^{(n)}=-\frac12\left[
 \overline F(k_1,k_2;P)+\overline F(k_2,k_1;P)\right].
 \label{eq:main-h-from-F}
\end{equation}
Equation~\eqref{eq:action-finite-difference} evaluates these two terms as
$-2\mathscr H_P(\beta)$ and $-2\mathscr H_P(\alpha)$, respectively.
Reflection of the box, $t_a\mapsto x_a-t_a$, maps one slice to the other and
therefore implies
$\mathscr H_P(\alpha)=\mathscr H_P(\beta)$. Hence
\begin{equation}
 h_{12}^{(n)}=2H_n
 =2(n-3)!\operatorname{Vol}(\mathcal W_n).
 \label{eq:contact-theorem}
\end{equation}
 Thus the Hydrotope geometry is already encoded in the contact interaction. As a simple illustration, consider
\begin{equation}
 (k_1,k_2,k_3,k_4)=(-7,-5,3,9).
 \label{eq:four-point-example-data}
\end{equation}
Then $\beta=5$ and
\begin{equation}
 \begin{aligned}
 \mathcal W_4&=\{0\leq t_3\leq3,\ 0\leq t_4\leq9,\ t_3+t_4=5\},\\
 H_4&=\operatorname{Vol}(\mathcal W_4)
 =5-\tp{5-3}-\tp{5-9}=3,\\
h_{12}^{(4)}&=7+5-|-7+3|-|-7+9|=6=2H_4.
 \end{aligned}
 \label{eq:four-point-contact-example}
\end{equation}
This example imposes only spatial-momentum conservation and probes the
independent-frequency contact kernel, rather than a physical resonant
process. Geometrically, $h_{12}^{(4)}$ is twice the normalized length of the
line segment obtained by slicing the rectangle. A derivation of
Eq.~\eqref{eq:action-finite-difference} is given in the Supplemental Material.

The denominator cancellation is not restricted to this sign sector. For any marked pair $i,j$, the fixed-pair coefficient admits a finite oriented box-sum representation,
\begin{equation}
 \begin{aligned}
 h_{ij}^{(n)}
 =\sum_{\phi\in\mathcal F_{ij}}\mu_\phi
   \prod_{r=1}^{n-3}L_{\phi,r}=\sum_{\phi\in\mathcal F_{ij}}\mu_\phi\,
   \operatorname{Vol}\!\left(\prod_{r=1}^{n-3}[0,L_{\phi,r}]\right).
 \end{aligned}
 \label{eq:flag-volume}
\end{equation}
Here $\phi$ denotes an ordered flag of spectator blocks,
$L_{\phi,r}$ the corresponding absolute prefix momenta, and $\mu_\phi$ the signed incidence weight. Their explicit definitions and the pole-cancellation proof are given in Eqs.~\smeqref{eq:app-L-definition}--\smeqref{eq:app-explicit-flag-volume}. For generic momenta, Eq.~\eqref{eq:flag-volume} defines an oriented chain of boxes rather than the volume of a single convex body, with Eq.~\eqref{eq:contact-theorem} arising as the Hydrotope specialization. Thus the action-level geometric structure extends beyond the two-minus sector; what is special there is that the oriented chain collapses to a single positive cut-box volume.

\textit{Complete two-minus tree sum.}---Having uncovered the geometry encoded in the contact vertex, we now study the complete on-shell amplitude in the two-minus sector to see how this structure is selected by the full tree sum and gives rise to the Hydrotope.

We now place each external leg on the dispersion relation while continuing to postpone overall frequency conservation. In the all-incoming convention, $w_i$ denotes the signed frequency, whereas the superscript $--+\cdots+$ refers to the signs of the spatial momenta, $\sigma_i:=\operatorname{sgn}(k_i)$, rather than to helicities or frequency
signs:
\begin{equation}
 \begin{gathered}
 k_i=\sigma_i w_i^2,\qquad \sum_i k_i=0,\\
 \sigma_{1,2}=-1,\qquad\sigma_a=+1\quad(a\in P).
 \end{gathered}
 \label{eq:on-shell}
\end{equation}

To reorganize the tree diagrams, we use the exact one-minus current~\cite{Ussembayev,Hydrotope},
\begin{equation}
 J(B)=k_{B}^{|B|-1}.
 \label{eq:rooted-tree-sum}
\end{equation}
Here $B$ is the external label set of a rooted component and the current includes its off-shell root propagator. After adjoining the root leg of momentum $-k_B$, every component appearing below has exactly one negative-momentum leg; Eq.~\eqref{eq:rooted-tree-sum} sums all of its rooted tree topologies into the displayed monomial.

A key observation provides an alternative organization of the
Berends--Giele tree sum in the two-minus sector. Every generic two-minus tree contains a unique distinguished vertex. Along the path from leg $1$ to leg $2$, the momentum starts at $k_1<0$ and increases whenever a plus-only branch is crossed, reaching $-k_2>0$ at the other end. It therefore changes sign exactly once, at the unique vertex where the two minus branches meet.

Cutting at this vertex, as shown in Fig.~\ref{fig:central-vertex}, decomposes the tree into rooted components. The components containing legs 1 and 2 are distinct and carry negative total momentum, while all others carry positive
total momentum. Equation~\eqref{eq:rooted-tree-sum} sums the internal topology of each component into a current $J(C)$. Conversely, gluing any admissible collection of these currents to the central contact vertex reconstructs a unique two-minus tree.

Let $\Adm_r(N)$ denote the partitions of $N$ into $r$ blocks such that the blocks containing $1$ and $2$ are distinct and both carry negative total momentum. For each block $C$, define
$k_C:=\sum_{i\in C}k_i$, $w_C:=\sum_{i\in C}w_i$, and
$p_C:=(k_C,w_C)$. The complete tree amplitude is then
\begin{equation}
 A_n^{(--+\cdots+)}
 =\sum_{r=3}^{n}\sum_{\pi\in\Adm_r(N)}
 \cV_r\bigl((p_C)_{C\in\pi}\bigr)
 \prod_{C\in\pi}k_C^{|C|-1}.
 \label{eq:central-tree-sum}
\end{equation}
The composite momenta $p_C$ are generically off shell,
$w_C^2\ne|k_C|$. At four points, this organization is completely explicit. Using vertical bars to separate the blocks of a partition, the contact diagram and the $12|34$ exchange channel correspond to
\begin{equation}
 \pi_4=1|2|3|4,
 \qquad
 \pi_{12|34}=1|2|34.
\end{equation}
A mixed channel has a unique admissible representative,
\begin{equation}
 \pi_{13|24}=
 \begin{cases}
 13|2|4,&k_{13}<0,\\
 1|3|24,&k_{24}<0,
 \end{cases}
 \qquad
 k_{13}=-k_{24},
\end{equation}
with $\pi_{14|23}$ obtained by $3\leftrightarrow4$. Replacing each two-leg block by its exact current, $J(ij)=k_i+k_j$, reproduces the corresponding exchange diagram. Thus Eq.~\eqref{eq:central-tree-sum} generates the contact diagram and all three four-point exchange diagrams exactly once.

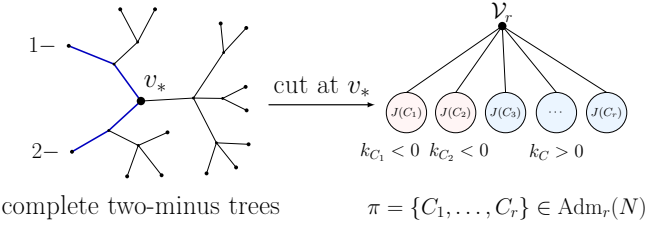
\begin{figure}[t]
  \definecolor{positivefill}{RGB}{235,245,255}
  \centering
  \resizebox{\columnwidth}{!}{%
  \begin{tikzpicture}[
    x=0.88cm,
    y=1cm,
    >=latex,
    line cap=round,
    line join=round,
    blackedge/.style={draw=black,line width=0.72pt},
    blueedge/.style={draw=blue!75!black,line width=1.20pt},
    vertex/.style={circle,fill=black,inner sep=0pt,minimum size=2.2pt},
    endpoint/.style={circle,fill=black,inner sep=0pt,minimum size=3.0pt},
    current/.style={
      circle,
      draw=black,
      line width=0.72pt,
      minimum size=1.08cm,
      inner sep=0pt,
      font=\small
    }
  ]

  \coordinate (one) at (0.92,5.05);
  \coordinate (up) at (2.30,4.56);
  \coordinate (vstar) at (3.12,3.57);
  \coordinate (low) at (2.14,2.78);
  \coordinate (two) at (1.02,2.21);
  \draw[blueedge] (one)--(up)--(vstar)--(low)--(two);

  \coordinate (uy) at (3.02,5.15);
  \draw[blackedge] (up)--(uy);
  \draw[blackedge] (uy)--(2.49,6.03);
  \draw[blackedge] (uy)--(3.55,6.03);

  \coordinate (ly) at (3.04,2.38);
  \draw[blackedge] (low)--(ly);
  \draw[blackedge] (ly)--(2.70,1.55);
  \draw[blackedge] (ly)--(3.75,1.58);
  \draw[blackedge] (ly)--(3.93,2.73);

  \coordinate (main) at (4.72,3.65);
  \coordinate (upperbranch) at (5.64,4.87);
  \coordinate (rightbranch) at (5.61,3.64);
  \coordinate (lowerbranch) at (5.42,2.45);
  \draw[blackedge] (vstar)--(main);
  \draw[blackedge] (main)--(upperbranch);
  \draw[blackedge] (upperbranch)--(5.55,5.65);
  \draw[blackedge] (upperbranch)--(6.33,5.53);
  \draw[blackedge] (main)--(rightbranch);
  \draw[blackedge] (rightbranch)--(6.31,3.95);
  \draw[blackedge] (rightbranch)--(6.34,3.32);
  \draw[blackedge] (main)--(lowerbranch);
  \draw[blackedge] (lowerbranch)--(5.12,1.63);
  \draw[blackedge] (lowerbranch)--(6.02,1.79);
  \draw[blackedge] (lowerbranch)--(6.40,2.41);

  \foreach \p in {up,low,uy,ly,main,upperbranch,rightbranch,lowerbranch}
    \node[vertex] at (\p) {};
  \foreach \p in {one,two}
    \node[endpoint] at (\p) {};
  \foreach \x/\y in {
    2.49/6.03,3.55/6.03,
    2.70/1.55,3.75/1.58,3.93/2.73,
    5.55/5.65,6.33/5.53,
    6.31/3.95,6.34/3.32,
    5.12/1.63,6.02/1.79,6.40/2.41
  } \node[endpoint] at (\x,\y) {};
  \node[circle,fill=black,inner sep=0pt,minimum size=6.4pt]
    at (vstar) {};

  \node[anchor=east,font=\LARGE] at (0.68,5.05) {$1-$};
  \node[anchor=east,font=\LARGE] at (0.78,2.21) {$2-$};
  \node[anchor=south west,font=\huge] at (3.08,3.71) {$v_*$};
  \node[font=\huge,anchor=base] at (3.12,0.55)
    {complete two-minus trees};

  \draw[->,line width=0.92pt] (7.02,3.47)--(10.25,3.47);
  \node[font=\huge] at (8.64,3.95)
    {cut at $v_*$};

  \coordinate (central) at (14.13,5.58);
  \node[font=\LARGE,anchor=south] at (central) {$\cV_r$};

  \foreach \x in {11.21,12.71,14.23,15.78,17.31}
    \draw[blackedge] (central)--(\x,3.80);

  \node[current,fill=red!4]  (c1) at (11.21,3.26) {$J(C_1)$};
  \node[current,fill=red!4]  (c2) at (12.71,3.26) {$J(C_2)$};
  \node[current,fill=positivefill] (c3) at (14.23,3.26) {$J(C_3)$};
  \node[current,fill=positivefill] (cd) at (15.78,3.26) {$\cdots$};
  \node[current,fill=positivefill] (cr) at (17.31,3.26) {$J(C_r)$};
  \node[circle,fill=black,inner sep=0pt,minimum size=6.4pt]
    at (central) {};

  \node[font=\Large] at (10.71,2.20) {$k_{C_1}<0$};
  \node[font=\Large] at (12.81,2.20) {$k_{C_2}<0$};
  \node[font=\Large] at (15.78,2.20) {$k_C>0$};

  \node[font=\LARGE,anchor=base] at (14.28,0.55)
    {$\pi=\{C_1,\ldots,C_r\}\in\Adm_r(N)$};

  \end{tikzpicture}%
  }
  \caption{Central-vertex reorganization of the complete two-minus tree
  sum. The thick path joins the two minus legs. Cutting at its unique
  sign-changing vertex $v_*$ replaces each rooted component by its exact
  current $J(C)=k_C^{|C|-1}$.}
  \label{fig:central-vertex}
\end{figure}

This partition representation also provides a natural way to resolve the complete tree sum into independent frequency-pair structures. Expanding the block frequencies $w_C$ in Eq.~\eqref{eq:central-tree-sum} in terms of the
external frequencies defines
\begin{equation}
A_n^{(--+\cdots+)}=\sum_{i<j}w_iw_jG_{ij}.
\label{eq:frequency-coefficients}
\end{equation}
Here $G_{ij}$ is defined before imposing frequency conservation. This prescription is essential, since terms proportional to $\sum_i w_i$ vanish upon imposing frequency conservation but can shift the individual coefficients $G_{ij}$. The resulting $G_{ij}$ are canonical representatives of the action-based decomposition rather than independent observables. Equation~\eqref{eq:central-tree-sum} has already absorbed all subsidiary tree topology into exact currents, leaving only the admissibility condition on the central partition.

The coefficients fall into three classes:
\begin{equation}
 \begin{gathered}
 G_{12}=(2^{n-1}-2)H_n,\,
 G_{1a}=G_{2a}=0,\, G_{ab}=2H_n,
 \end{gathered}
 \label{eq:three-identities}
\end{equation}
where $a,b\in P$ and $a\ne b$. These identities hold before imposing frequency conservation and completely determine the quadratic frequency polynomial. In particular, the mixed coefficients vanish directly in the complete tree sum, rather than as a consequence of $\sum_i w_i=0$.
\begin{figure}[!t]
  \definecolor{positivefill}{RGB}{235,245,255}
  \definecolor{negativefill}{RGB}{255,240,240}
  \definecolor{markedfill}{RGB}{232,248,236}
  \centering
  \begin{tikzpicture}[
    x=0.72cm,
    y=0.68cm,
    >=latex,
    line cap=round,
    line join=round,
    edge/.style={draw=black,line width=0.72pt},
    vertex/.style={circle,fill=black,inner sep=0pt,minimum size=5.4pt},
    current/.style={
      circle,draw=black,line width=0.72pt,minimum size=0.80cm,
      inner sep=1pt,font=\scriptsize
    },
    minicurrent/.style={
      circle,draw=black,line width=0.68pt,minimum size=0.72cm,
      inner sep=0pt,font=\scriptsize
    },
    excluded/.style={draw=gray!75,dashed,line width=0.68pt,rounded corners=3pt}
  ]

  \node[font=\small] at (5.05,6.45)
    {\textbf{(a)}\quad mixed pair $G_{1a}$};

  \coordinate (vl) at (2.30,5.95);
  \node[vertex] at (vl) {};
  \node[font=\footnotesize,anchor=south] at (2.30,6.10) {$\mathcal V_r$};
  \node[current,fill=negativefill] (l1) at (0.62,4.55) {$J(C_1)$};
  \node[current,fill=negativefill] (l2) at (1.74,4.55) {$J(C_2)$};
  \node[current,fill=markedfill,draw=ForestGreen!70!black,line width=0.92pt]
    (la) at (2.86,4.55) {$a$};
  \node[current,fill=positivefill] (lu) at (3.98,4.55) {$\cdots$};
  \foreach \n in {l1,l2,la,lu} \draw[edge] (vl)--(\n);
  \node[font=\footnotesize,text=ForestGreen!55!black] at (2.30,3.66)
    {$a$ exposed at $\mathcal V_r$};
  \node[font=\Large,text=ForestGreen!55!black] at (0.06,5.58) {$+$};

  \coordinate (vr) at (7.25,5.95);
  \node[vertex] at (vr) {};
  
  \node[current,fill=negativefill] (r1) at (5.82,4.55) {$J(C_1)$};
  \node[current,fill=negativefill,draw=ForestGreen!70!black,line width=0.92pt,
    inner sep=0pt]
    (r2a) at (7.25,4.55)
    {\resizebox{0.98cm}{!}{$J(C_2')$}};
  \node[current,fill=positivefill] (ru) at (8.68,4.55) {$\cdots$};
  \foreach \n in {r1,r2a,ru} \draw[edge] (vr)--(\n);
  \node[font=\footnotesize,text=ForestGreen!55!black] at (7.25,3.52)
    {$C_2' := C_2 \cup \{a\}$};
  \node[font=\Large,text=BrickRed] at (5.12,5.58) {$-$};
  \node[font=\Large] at (9.85,5.10) {$=0$};

  \node[font=\footnotesize] at (5.05,3.08)
    {$[G(\xi)-G(0)]\Psi_\rho=0
      \quad\Longrightarrow\quad G_{1a}=0
      \quad (1\leftrightarrow2:\ G_{2a}=0)$};

  \draw[gray!45,line width=0.55pt] (0.00,2.62)--(10.15,2.62);

  \begin{scope}[yshift=4.5mm]
  \node[font=\small] at (5.05,1.53)
    {\textbf{(b)}\quad plus pair $G_{ab}$};

  \node[font=\footnotesize,anchor=west] at (0.16,1.06)
    {$\mathcal U_{ab}=0$};
  \coordinate (vu) at (2.18,0.62);
  \node[vertex] at (vu) {};
  \node[minicurrent,fill=gray!10,draw=gray!80!black] (u1)
    at (0.64,-0.48) {$C_1^{\pm}$};
  \node[minicurrent,fill=gray!10,draw=gray!80!black] (u2)
    at (1.66,-0.48) {$C_2^{\pm}$};
  \node[minicurrent,fill=markedfill,draw=ForestGreen!70!black] (ua)
    at (2.70,-0.48) {$C_a$};
  \node[minicurrent,fill=markedfill,draw=ForestGreen!70!black] (ub)
    at (3.72,-0.48) {$C_b$};
  \foreach \n in {u1,u2,ua,ub} \draw[edge] (vu)--(\n);
  \node[font=\scriptsize] at (2.18,-1.30) {all dressed partitions};

  \draw[excluded] (4.98,1.16) rectangle (9.81,-1.52);
  \node[font=\footnotesize,anchor=north west] at (5.28,1.06)
    {$-\,\mathcal S_{12}$};
  \coordinate (vs) at (7.30,0.62);
  \node[vertex] at (vs) {};
  \node[minicurrent,fill=negativefill,minimum size=0.82cm] (s12)
    at (6.18,-0.48) {$C_{12}$};
  \node[minicurrent,fill=markedfill,draw=ForestGreen!70!black] (sa)
    at (7.30,-0.48) {$C_a$};
  \node[minicurrent,fill=markedfill,draw=ForestGreen!70!black] (sb)
    at (8.42,-0.48) {$C_b$};
  \foreach \n in {s12,sa,sb} \draw[edge] (vs)--(\n);
  \node[font=\scriptsize] at (7.30,-1.30) {$1,2$ in the same block};

  \draw[excluded] (0.00,-1.68) rectangle (4.83,-4.65);
  \node[font=\footnotesize,anchor=north west] at (0.16,-1.80)
    {$-\,\mathcal C_{\alpha|\beta}$};
  \coordinate (va) at (2.26,-2.30);
  \node[vertex] at (va) {};
  \node[minicurrent,fill=negativefill] (a1) at (0.68,-3.40) {$C_1^-$};
  \node[minicurrent,fill=positivefill] (a2) at (1.73,-3.40) {$C_2^+$};
  \node[minicurrent,fill=markedfill,draw=ForestGreen!70!black] (aa)
    at (2.78,-3.40) {$C_a$};
  \node[minicurrent,fill=markedfill,draw=ForestGreen!70!black] (ab)
    at (3.83,-3.40) {$C_b$};
  \foreach \n in {a1,a2,aa,ab} \draw[edge] (va)--(\n);
  \node[font=\scriptsize] at (2.42,-4.30)
    {only the $1$-block is negative};

  \draw[excluded] (4.98,-1.68) rectangle (9.81,-4.65);
  \node[font=\footnotesize,anchor=north west] at (5.14,-1.80)
    {$-\,\mathcal C_{\beta|\alpha}$};
  \coordinate (vb) at (7.28,-2.30);
  \node[vertex] at (vb) {};
  \node[minicurrent,fill=positivefill] (b1) at (5.70,-3.40) {$C_1^+$};
  \node[minicurrent,fill=negativefill] (b2) at (6.75,-3.40) {$C_2^-$};
  \node[minicurrent,fill=markedfill,draw=ForestGreen!70!black] (ba)
    at (7.80,-3.40) {$C_a$};
  \node[minicurrent,fill=markedfill,draw=ForestGreen!70!black] (bb)
    at (8.85,-3.40) {$C_b$};
  \foreach \n in {b1,b2,ba,bb} \draw[edge] (vb)--(\n);
  \node[font=\scriptsize] at (7.40,-4.30)
    {only the $2$-block is negative};

\end{scope}
  \end{tikzpicture}

  \caption{Frequency-pair identities after the central cut.
(a) For the mixed pair $G_{1a}$, the marked plus leg $a$ is either exposed at the central vertex or absorbed into the leg-$2$ dressed current, $J(C_2')$, with $C_2'=C_2\cup \{a\}$. The exact harmonic-mode identity cancels these two fully dressed sectors,
giving $G_{1a}=0$; exchanging $1\leftrightarrow2$ gives $G_{2a}=0$.
(b) For the plus pair $G_{ab}$, $C_a$ and $C_b$ denote dressed blocks
containing the marked labels $a$ and $b$. The unrestricted harmonic
completion gives $\mathcal U_{ab}=0$. Removing the three excluded classes
leaves
$G_{ab}=\mathcal U_{ab}-\mathcal S_{12}
-\mathcal C_{\alpha|\beta}-\mathcal C_{\beta|\alpha}=2H_n$.
Grey dashed sectors belong to the unrestricted harmonic completion.}

  \label{fig:G1a/Gab}
  
\end{figure}
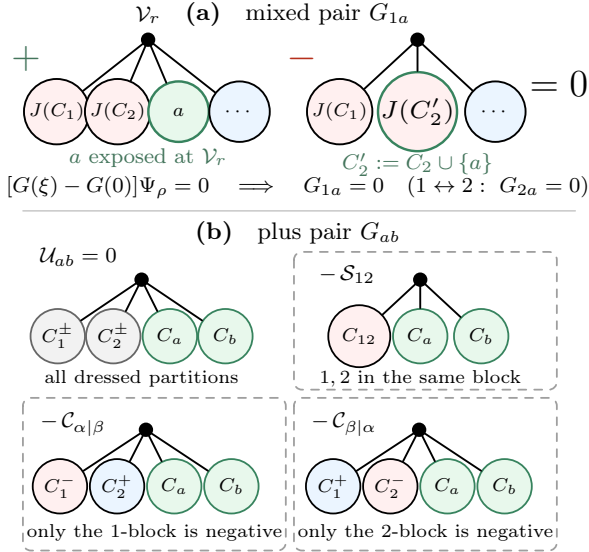
For the four-point kinematics of Eq.~\eqref{eq:four-point-example-data}, they reduce to
\begin{equation}
 (G_{12},G_{13},G_{14},G_{23},G_{24},G_{34})
 =(18,0,0,0,0,6),
\end{equation}
making explicit both the cancellation of the mixed coefficients and the enhancement of the minus–minus coefficient relative to its contact value.

The first identity can be refined by the number of blocks in the central partition. Let $G_{12}^{[s+2]}$ denote the contribution with $s$ plus-only blocks, and hence $s{+}2$ blocks in total. Then
\begin{equation}
 \sum_{s=1}^{n-2}\tau^sG_{12}^{[s+2]}
 =2\bigl[(1+\tau)^{n-2}-1\bigr]H_n,
 \label{eq:G12-generating-function}
\end{equation}
so that $G_{12}^{[s+2]}=2\binom{n-2}{s}H_n$. At fixed $s$, the rooted-current sum collapses to the same $H_n$, while the binomial coefficient counts the allowed distributions of the plus labels among the $s{+}2$ blocks. The labelled coefficient extraction is given in Eqs.~\smeqref{eq:app-fixed-R}--\smeqref{eq:app-G12} of the Supplemental Material.

The remaining two identities follow from an exact positive-momentum boundary mode. For $\rho>0$, define
\begin{equation}
 \Psi_\rho[\xi](x)=e^{\ii\rho x+\rho\xi(x)},\qquad
 G(\xi)\Psi_\rho[\xi]=-\ii\partial_x\Psi_\rho[\xi].
 \label{eq:positive-harmonic-mode}
\end{equation}
Expanding this exact boundary mode in $\xi$ yields contact identities at every multiplicity, which remain valid after replacing each rooted component by its exact current. 

For $G_{1a}$, the differentiated plus leg either enters the central vertex or lies in the block containing minus leg 2. Fix the block
containing minus leg 1, and let $A>0$ denote the total momentum carried by the remaining plus-only component. In the matrix element below, $\xi$ contains only positive-momentum modes from this component. Since $\rho>0$, every Fourier component of $\Psi_\rho[\xi]$ therefore has positive momentum, so
$G(0)=|\partial_x|$ acts on $\Psi_\rho[\xi]$ exactly as
$-\ii\partial_x$. Together with Eq.~\eqref{eq:positive-harmonic-mode}, this implies
\begin{equation}
 \left\langle e^{-\ii Ax},
 [G(\xi)-G(0)]\Psi_\rho[\xi]\right\rangle=0.
 \label{eq:mixed-trace-zero}
\end{equation}
Expanding Eq.~\eqref{eq:mixed-trace-zero} organizes the contributions into two fully dressed classes: the differentiated plus leg either attaches directly to the central vertex or belongs to the exact current on the leg-2 side. For each fixed block containing leg 1, summing the remaining labels over all allowed partitions and current dressings makes these two classes cancel exactly, as illustrated in Fig.~\ref{fig:G1a/Gab}. Exchanging legs 1 and 2 then gives $G_{2a}=0$.

This cancellation becomes local to the distinguished vertex only after the adjacent rooted sector has been resummed into its exact current. It neither uses frequency conservation nor mixes different frequency-pair structures, and should therefore be understood as a cancellation between fully dressed sectors rather than between individual graphs.

For $G_{ab}$, we first relax the admissibility condition and sum over all partitions, allowing the two minus legs to lie in the same block or in blocks that are not both of negative total momentum. Self-adjointness of $G(\xi)$, together with Eq.~\eqref{eq:positive-harmonic-mode}, then gives for $p,q>0$
\begin{equation}
 2\langle\Psi_p,G(\xi)\Psi_q\rangle
 =-\ii\int\dd x\,\partial_x(\Psi_p\Psi_q)=0.
 \label{eq:unrestricted-positive-zero}
\end{equation}

The total derivative vanishes for periodic waves or for fields decaying at spatial infinity, so Eq.~\eqref{eq:unrestricted-positive-zero} gives the unrestricted plus-pair sum. The physical tree sum excludes three disjoint classes: the two minus legs lie in the same block, or only the block containing leg 1 or leg 2 has negative total momentum. These exhaust the complement of $\Adm_r(N)$. Subtracting the excluded classes from the unrestricted sum reproduces the same inclusion--exclusion structure as Eq.~\eqref{eq:hydrotope},
\begin{equation}
 G_{ab}=\mathcal U_{ab}-\mathcal S_{12}
 -\mathcal C_{\alpha|\beta}-\mathcal C_{\beta|\alpha}=2H_n.
 \label{eq:plus-pair-subtraction}
\end{equation}
Here $\mathcal U_{ab}$ denotes the unrestricted sum, $\mathcal S_{12}$ the same-block contribution, and $\mathcal C_{\alpha|\beta}$ and $\mathcal C_{\beta|\alpha}$ the two one-negative-block contributions, as illustrated in Fig.~\ref{fig:G1a/Gab}. Their Laplace weights and the resulting
subtraction are given in Eqs.~\smeqref{eq:app-plus-Laplace-data}--\smeqref{eq:app-plus-plus} of the Supplemental Material.

Taken together, Eqs.~\eqref{eq:G12-generating-function},
\eqref{eq:mixed-trace-zero}, and \eqref{eq:plus-pair-subtraction} show that the simplicity of the full tree amplitude is not generated from scratch by the diagram sum. Rather, the tree sum reorganizes and cancels structures already present in the contact interaction, reducing them in the two-minus sector to
the single function $H_n$. In other sign sectors, the surviving combination need not admit a single-Hydrotope representation, but the action-level structure suggests an organization considerably simpler than the individual tree diagrams.

\textit{On-shell amplitude and cancellation.}---Having resolved the complete tree sum into its three independent frequency-pair coefficients, we now impose frequency conservation to recover the physical on-shell amplitude. Together with Eq.~\eqref{eq:on-shell}, this gives
\begin{equation}
 \sum_{3\leq a<b\leq n}w_aw_b
 =\frac12\left[\left(\sum_{a\in P}w_a\right)^2
 -\sum_{a\in P}w_a^2\right]
 =w_1w_2,
 \label{eq:positive-pair-sum}
\end{equation}
where $\sum_{a\in P}w_a=-(w_1+w_2)$ and $\sum_{a\in P}w_a^2=w_1^2+w_2^2$. Combining Eqs.~\eqref{eq:three-identities} and \eqref{eq:positive-pair-sum} then yields
\begin{equation}
 A_n^{(--+\cdots+)}=2^{n-1}w_1w_2H_n.
 \label{eq:amplitude}
\end{equation}
Thus the Hydrotope amplitude follows from the three coefficient identities, rather than being used as an input to derive them.

\textit{Conclusion and outlook.}---We have traced the Hydrotope geometry directly to the water-wave action. For any fixed pair of legs, the corresponding contact coefficient admits a denominator-free representation as a finite oriented sum of box volumes, reducing to the ordinary Hydrotope whenever the momentum-sign assignment is compatible with the chosen pair. In the complete two-minus amplitude, the unique central-vertex decomposition glues these contact structures to exact one-minus currents. The three identities in Eq.~\eqref{eq:three-identities} then expose the mechanism by which the sum over trees cancels all mixed-sign coefficients and reorganizes the surviving terms into the Hydrotope amplitude \eqref{eq:amplitude}. The remarkable simplicity of the final result is therefore not an accidental property of the amplitude, but descends from geometric structures already encoded in the action.

These results point toward a broader organizing principle for water-wave amplitudes. Beyond the two-minus sector, a general amplitude need not collapse to the volume of a single Hydrotope, but the action-level geometry may nevertheless continue to organize its structure and enforce nontrivial cancellations.\footnote{The two-minus sector has no physical factorization poles. A nonvanishing factorization channel would require at least two legs of each momentum sign on both sides of the cut, which is impossible when there are only two external minus legs.} Further evidence is provided by a soft theorem valid for arbitrary momentum-sign configurations of complete tree amplitudes, whose proof and applications will be presented elsewhere. Combined with factorization, these soft constraints furnish a recursive framework for higher-minus amplitudes and a systematic route to uncovering further cancellations and geometric structures.

\textit{Acknowledgments.}---OpenAI's GPT-5.6 Sol model, accessed through Codex, assisted with derivations, computational checks, code development, and manuscript preparation; all AI-assisted material and scientific conclusions were reviewed and verified by the authors, who take full responsibility for the content. Q.C. is supported by the Westlake Fellows Program at Westlake University. S.H. is supported by the National Natural Science Foundation of China under Grant Nos. 12225510 and 12447101, and by the New Cornerstone Science Foundation. J.J. and Q.L. thank Westlake University for its hospitality during the completion of this work.

\bibliographystyle{apsrev4-2}
\bibliography{Refs}
\fi

\ifdefined\HYDROMAINONLY
  \def\hydronext{\end{document}}
\else
  \def\hydronext{}
\fi
\hydronext

\ifdefined\HYDROSUPPLEMENT
  \onecolumngrid
  \appendix
\else
  \clearpage
  \onecolumngrid
  \appendix
\fi

\section{Fixed-pair contact coefficients}
\label{sec:app-fixed-pair}

This section proves the flag formula in Eq.~\maineqref{eq:flag-volume}.  We use
the notation of the Letter throughout: for a set of labels $A$,
$k_A:=\sum_{a\in A}k_a$, and $h_{ij}^{(n)}$ is the coefficient multiplying
$w_iw_j$ in Eq.~\maineqref{eq:contact-polarization}.  No additional sign factor
is to be attached to $h_{ij}^{(n)}$.

Fix the ordered marked pair $a|b$, where eventually $a=k_i$ and $b=k_j$,
and let $R$ be the set of the remaining labels.  Put $r=|R|$.  Every kernel
below obeys its local spatial-momentum conservation law; in particular,
$a+b+k_R=0$.  We first work in an open chamber in which

\begin{equation}
 b+k_U\neq0\qquad(U\subseteq R).
 \label{eq:app-open-chamber}
\end{equation}

The result on a wall is the one-sided limit of the complete expression, not
the result of assigning $\operatorname{sgn}(0)$ to its individual terms.

\subsection{Cancellation of the apparent kernel poles}

We first connect the notation below directly to the action.  Expanding
Eq.~\maineqref{eq:direct-action}, let
$E_{m+2}(a,b;y_1,\ldots,y_m)$ denote the ordered kernel with momenta $a,b$
in the two $\chi$ slots and the remaining fields in the displayed order.
With $Y_q:=y_1+\cdots+y_q$, the Dirichlet--Neumann recursion is

\begin{align}
 E_{m+2}(a,b;y_1,\ldots,y_m)
 ={}&\frac{|b|^{m-1}}{m!}E_3(a,b)-\sum_{q=1}^{m-1}\frac{|b|^q}{q!}
 E_{m-q+2}(a,b+Y_q;y_{q+1},\ldots,y_m).
 \label{eq:app-ordered-E}
\end{align}

We use a sum, rather than an average, over spectator orders,

\begin{equation}
 \overline E(a,b;R):=
 \sum_{\pi\in S_R}
 E_{r+2}(a,b;k_{\pi(1)},\ldots,k_{\pi(r)}),
 \qquad r=|R|,
 \label{eq:app-symmetrized-E}
\end{equation}
so that each labelled ordered partition below has unit multiplicity. 

Let $\overline F(a,b;R)$ denote the corresponding spectator-symmetrized
directed kernel from the inverse Dirichlet--Neumann expansion. We work on the nonzero Fourier modes, with the spatial zero mode projected
out.  Defining
\begin{equation}
 \chi:=\partial_x^{-1}\dot\xi,
 \qquad
 \dot\xi=\partial_x\chi,
 \label{eq:app-chi-definition}
\end{equation}
and using
\begin{equation}
 G(\xi)^{-1}
 =-\partial_x^{-1}G(\xi)\partial_x^{-1},
 \qquad
 \bigl(\partial_x^{-1}\bigr)^\dagger
 =-\partial_x^{-1},
 \label{eq:app-harmonic-operator-identity}
\end{equation}
we obtain
\begin{align}
 \left\langle\dot\xi,G(\xi)^{-1}\dot\xi\right\rangle
 &=
 -\left\langle
 \dot\xi,\partial_x^{-1}G(\xi)\partial_x^{-1}\dot\xi
 \right\rangle
 \nonumber\\
 &=
 \left\langle\chi,G(\xi)\chi\right\rangle .
 \label{eq:app-quadratic-harmonic-duality}
\end{align}
This is the quadratic-form version of
Eq.~\maineqref{eq:harmonic-duality}.  It may be understood with periodic
boundary conditions or, equivalently, with a standard wave-packet
regularization.

The equality of quadratic forms first determines only the
endpoint-symmetrized kernel.  To recover the directed relation, note that
$G(\xi)$ and $G(\xi)^{-1}$ are self-adjoint order by order in $\xi$.
Exchanging the two marked endpoints reverses the spectator ordering; since
$\overline E$ and $\overline F$ contain the sum over all spectator
permutations, this gives
\begin{equation}
 \overline E(a,b;R)=\overline E(b,a;R),
 \qquad
 \overline F(a,b;R)=\overline F(b,a;R).
 \label{eq:app-endpoint-exchange}
\end{equation}
Before using these identities, coefficient extraction from
Eq.~\eqref{eq:app-quadratic-harmonic-duality}, with the kernel
normalizations used here, gives
\begin{equation}
 \overline F(a,b;R)+\overline F(b,a;R)
 =
 \frac{2}{ab}
 \left[
 \overline E(a,b;R)+\overline E(b,a;R)
 \right].
 \label{eq:app-EF-symmetrized-bridge}
\end{equation}
The two factors of $\partial_x^{-1}$ contribute
$(\mathrm{i}a)^{-1}(\mathrm{i}b)^{-1}$, whose minus sign is cancelled by
the overall minus sign in
Eq.~\maineqref{eq:harmonic-duality}.  Combining
Eqs.~\eqref{eq:app-endpoint-exchange} and
\eqref{eq:app-EF-symmetrized-bridge} therefore yields the directed identity
\begin{equation}
 \boxed{
 \overline F(a,b;R)
 =\frac{2\overline E(a,b;R)}{ab}.}
 \label{eq:app-EF-bridge}
\end{equation}
At cubic order this normalization is confirmed by
\begin{equation}
 \overline F_3(a,b)
 =\frac{2E_3(a,b)}{ab}
 =-\left[
 1+\operatorname{sgn}(a)\operatorname{sgn}(b)
 \right].
\end{equation}

Only the endpoint-symmetrized combination enters the physical unordered
frequency-pair coefficient:
\begin{equation}
 \boxed{
 \begin{aligned}
 h_{ij}^{(n)}
 &=-\frac1{k_i k_j}
 \left[
 \overline E(k_i,k_j;R)
 +\overline E(k_j,k_i;R)
 \right]
 \\
 &=-\frac12
 \left[
 \overline F(k_i,k_j;R)
 +\overline F(k_j,k_i;R)
 \right].
 \end{aligned}}
 \label{eq:app-action-kernel-bridge}
\end{equation}

This also fixes all factorial conventions in what follows. Symmetrizing
Eq.~\eqref{eq:app-ordered-E} gives the first recursion below. Independently,
expanding $G(\xi)^{-1}$ and using
$G(\xi)G(\xi)^{-1}=1$ gives the second; Eq.~\eqref{eq:app-EF-bridge}
states their equality after harmonic duality:

\begin{align}
 \overline E(a,b;R)
 ={}&|b|^{r-1}E_3(a,b)
 -\sum_{\varnothing\ne U\subsetneq R}
 |b|^{|U|}\overline E(a,b+k_U;R\setminus U),
 \label{eq:app-E-recursion}\\
 \overline F(a,b;R)
 ={}&\frac1{|b|}\left[
 \frac{2\overline E(a,b;R)}{|a|}
 -\sum_{\varnothing\ne U\subsetneq R}
 2\overline E(-(b+k_U),b;U)
 \overline F(a,b+k_U;R\setminus U)\right],
 \label{eq:app-F-recursion}
\end{align}

with cubic seed

\begin{equation}
 E_3(a,b)=-\frac12\bigl(|a||b|+ab\bigr).
 \label{eq:app-E3}
\end{equation}

Thus Eqs.~\eqref{eq:app-EF-bridge} and \eqref{eq:app-F-recursion} implement
harmonic duality and inverse composition, respectively.

The denominators in Eq.~\eqref{eq:app-F-recursion} are only apparent.  Define
$\widehat E$ by

\begin{equation}
 \boxed{
 2\overline E(a,b;R)=|a||b|\widehat E(a,b;R).}
 \label{eq:app-endpoint-factor}
\end{equation}

For $r=1$ this follows directly from Eq.~\eqref{eq:app-E3}, since

\begin{equation}
 \widehat E_3(a,b)
 =-\bigl[1+\operatorname{sgn}(a)\operatorname{sgn}(b)\bigr].
 \label{eq:app-Ehat3}
\end{equation}

Assuming Eq.~\eqref{eq:app-endpoint-factor} for smaller spectator sets and
substituting it into Eq.~\eqref{eq:app-E-recursion} gives

\begin{align}
 \widehat E(a,b;R)
 ={}&|b|^{r-1}\widehat E_3(a,b)
 -\sum_{\varnothing\ne U\subsetneq R}
 |b|^{|U|-1}|b+k_U|
 \widehat E(a,b+k_U;R\setminus U).
 \label{eq:app-Ehat-recursion}
\end{align}

Because $U$ is nonempty, every exponent on the right is nonnegative.  This
proves Eq.~\eqref{eq:app-endpoint-factor} by induction.  The same
substitution in Eq.~\eqref{eq:app-F-recursion} removes its remaining
denominators and yields

\begin{equation}
 \boxed{
 \begin{aligned}
 \overline F(a,b;R)={}&\widehat E(a,b;R)-\sum_{\varnothing\ne U\subsetneq R}
 |b+k_U|\widehat E(-(b+k_U),b;U)
 \overline F(a,b+k_U;R\setminus U).
 \end{aligned}}
 \label{eq:app-F-regular}
\end{equation}

Thus the directed contact kernel is a homogeneous piecewise polynomial of
degree $r-1=n-3$.

\subsection{Closed flag formula}

For an ordered partition
$\pi=(B_1,\ldots,B_\ell)\in\OP(U)$ of a nonempty subset $U\subseteq R$,
define

\begin{equation}
 U_t:=B_1\sqcup\cdots\sqcup B_t,
 \qquad P_t:=b+k_{U_t},
 \qquad P_0:=b.
 \label{eq:app-prefixes}
\end{equation}

Iterating Eq.~\eqref{eq:app-Ehat-recursion} gives

\begin{align}
 \widehat E(a,b;R)
 ={}&\sum_{(B_1,\ldots,B_\ell)\in\OP(R)}
 (-1)^{\ell-1}|P_0|^{|B_1|-1}
 \left(\prod_{t=2}^{\ell}|P_{t-1}|^{|B_t|}\right)
 \widehat E_3(a,P_{\ell-1}).
 \label{eq:app-Ehat-ordered}
\end{align}

Before collecting refinements, iteration of
Eq.~\eqref{eq:app-F-regular} gives the intermediate chain

\begin{align}
 \overline F(a,b;R)
 =\sum_{\pi=(B_1,\ldots,B_\ell)\in\OP(R)}T_\pi,\quad
 T_\pi
 =(-1)^{\ell-1}
 \left(\prod_{t=1}^{\ell-1}|P_t|\right)
 \widehat E(a,P_{\ell-1};B_\ell)
 \prod_{t=1}^{\ell-1}
 \widehat E(-P_t,P_{t-1};B_t).
 \label{eq:app-F-chain}
\end{align}

Indeed, $B_1$ is the unique subset chosen in the first convolution term of
Eq.~\eqref{eq:app-F-regular}; iteration fixes all later blocks.  Thus no
factorial or block-order multiplicity is hidden in
Eq.~\eqref{eq:app-F-chain}.  Insert
Eq.~\eqref{eq:app-Ehat-ordered} in each factor and group histories with the
same ordered blocks.  A final monomial is labelled by

\begin{equation}
 \phi=(B_1|\cdots|B_\ell;T),
 \qquad
 T:=R\setminus U.
 \label{eq:app-flag-atom}
\end{equation}

For a fixed final atom
$\phi=(B_1|\cdots|B_\ell;T)$, let
$I_\phi:=\{1,\ldots,\ell-1\}$ denote the internal boundaries
between consecutive blocks.  After substituting the expansion of
$\widehat E$ into the intermediate $F$ chain, each recursive history
determines nested boundary sets $D\subseteq C\subseteq I_\phi$: $D$ contains
the boundaries exposed by the outer $F$ chain, while $C\setminus D$ contains
those created inside the $\widehat E$ factors.  For fixed $D$, toggling one
boundary in $I_\phi\setminus D$ changes one recursion sign but not the
associated prefix momentum, because the two adjacent blocks recombine to
the same prefix.  Hence all choices of $C$ have the same
absolute-momentum monomial and their coefficient factorizes as
\begin{equation}
 \sum_{D\subseteq C\subseteq I_\phi}
 (-1)^{|C|-|D|}
 =
 (1-1)^{|I_\phi\setminus D|}.
 \label{eq:s-boundary-cancellation}
\end{equation}
It vanishes unless $D=I_\phi$.  Thus every history that leaves an
internal boundary unresolved cancels.  In the surviving term, all
internal prefixes occur through their absolute values, while only
the two endpoint signs
$\operatorname{sgn}(P_0)$ and $\operatorname{sgn}(P_\ell)$ remain.
Together with the base term $-|b|^{r-1}$, this gives

\begin{equation}
 \boxed{\begin{aligned}
 \overline F(a,b;R)={}&-|b|^{r-1}
 +\sum_{\varnothing\ne U\subseteq R}
 \sum_{(B_1,\ldots,B_\ell)\in\OP(U)}
 (-1)^{\ell+1}
 \operatorname{sgn}(P_0)\operatorname{sgn}(P_\ell)
\times |P_0|^{|B_1|-1}
 \left(\prod_{t=2}^{\ell}|P_{t-1}|^{|B_t|}\right)
 |P_\ell|^{|T|}.
 \end{aligned}}
 \label{eq:app-F-flag}
\end{equation}

For later reference we repeat the physical, unordered pair coefficient,

\begin{equation}
 \boxed{
 h_{ij}^{(n)}=-\frac12\left[
 \overline F(k_i,k_j;R)+\overline F(k_j,k_i;R)\right],
 \qquad R=N\setminus\{i,j\}.}
 \label{eq:app-h-from-F}
\end{equation}

This is Eq.~\eqref{eq:app-action-kernel-bridge}; it makes
$h_{ij}^{(n)}=h_{ji}^{(n)}$ manifest.

We now spell out the lengths and weights used in Eq.~\maineqref{eq:flag-volume}.
Set $d=n-3=r-1$.  Each directed order $a|b$ contains a base atom
$\phi_0^{a|b}$ with $\mathcal L_{\phi_0^{a|b}}=\{|b|^{\times d}\},\quad\mu_{\phi_0^{a|b}}=\frac12$.

Here $X^{\times q}$ means $q$ copies of $X$; zero copies contribute no
factor.  For the nontrivial atom in Eq.~\eqref{eq:app-flag-atom}, define the
multiset

\begin{equation}
 \boxed{
 \mathcal L_\phi=
 \left\{
 |P_0|^{\times(|B_1|-1)},
 |P_1|^{\times|B_2|},\ldots,
 |P_{\ell-1}|^{\times|B_\ell|},
 |P_\ell|^{\times|T|}
 \right\},}
 \label{eq:app-L-definition}
\end{equation}

Indeed,
$(|B_1|-1)+\sum_{t=2}^{\ell}|B_t|+|T|
=|U|-1+|T|=r-1=d$,
so the multiset contains exactly $d$ lengths.  Its weight is
\begin{equation}
 \mu_\phi
 =
 \frac{1}{2}(-1)^\ell
 \operatorname{sgn}(P_0)\operatorname{sgn}(P_\ell).
 \label{eq:s-explicit-mu}
\end{equation}
Let $\mathcal F_{i|j}$ denote the collection consisting of the base atom
and all nontrivial atoms for the order $k_i|k_j$, and set
$\mathcal F_{ij}:=\mathcal F_{i|j}\sqcup\mathcal F_{j|i}$.
Listing the elements of $\mathcal L_\phi$ as
$L_{\phi,1},\ldots,L_{\phi,d}$ gives the promised explicit formula

\begin{equation}
 \boxed{
 h_{ij}^{(n)}
 =\sum_{\phi\in\mathcal F_{ij}}\mu_\phi
 \prod_{q=1}^{d}L_{\phi,q}
 =\sum_{\phi\in\mathcal F_{ij}}\mu_\phi
 \operatorname{Vol}\!\left(\prod_{q=1}^{d}[0,L_{\phi,q}]\right).}
 \label{eq:app-explicit-flag-volume}
\end{equation}

For $d=0$, the empty product is a point of volume one.  In mixed chambers
the weights in Eq.~\eqref{eq:s-explicit-mu} need not be positive; hence
Eq.~\eqref{eq:app-explicit-flag-volume} is in general an oriented sum of
boxes, rather than the volume of a single convex body.

\subsection{Low-point checks}

At three points, momentum conservation immediately gives

\begin{equation}
 \boxed{
 h_{ij}^{(3)}
 =1+\operatorname{sgn}(k_i)\operatorname{sgn}(k_j).}
 \label{eq:app-three-point}
\end{equation}

Thus, for nonzero conserved momenta, only the unique same-sign pair
contributes.  For example,
$k=(-3,-2,5)$ gives $(h_{12},h_{13},h_{23})=(2,0,0)$ and
$\cV_3=2w_1w_2$.  This is a check of the contact seed, not a real
three-wave resonance.

At four points, let $\{c,d\}=N\setminus\{i,j\}$.  The
one-dimensional flag sum becomes
\begin{equation}
 h_{ij}^{(4)}
 =
 \operatorname{sgn}(k_i)\operatorname{sgn}(k_j)
 \left(
 |k_i|+|k_j|-|k_i+k_c|-|k_i+k_d|
 \right).
 \label{eq:s-four-point-pair}
\end{equation}

For the conserved point $k=(-7,-5,3,9)$, the six coefficients are $(h_{12},h_{13},h_{14},h_{23},h_{24},h_{34})
 =(6,4,0,8,0,6)$.

The pairs $12$ and $34$ are pair-compatible and give $2H_4=6$.  The mixed
entries are signed flag continuations and must not be interpreted as ordinary
positive Hydrotope volumes.

\section{Reduction to the ordinary Hydrotope}
\label{sec:app-hydrotope}

We now prove Eq.~\maineqref{eq:contact-theorem}.  Following the notation of the
Letter, take

\begin{equation}
 k_1=-\alpha,\qquad k_2=-\beta,\qquad
 k_a=x_a>0\ (a\in P),\qquad
 P=\{3,\ldots,n\},\qquad \alpha\geq\beta>0.
 \label{eq:app-compatible-data}
\end{equation}

Momentum conservation gives $x_P:=\sum_{a\in P}x_a=\alpha+\beta$.  No
dispersion relation or frequency conservation is used in this section.

For an arbitrary slice position $u$, introduce the auxiliary truncated-power
function

\begin{equation}
 \mathscr H_P(u):=
 \sum_{A\subseteq P}(-1)^{|A|}\tp{u-x_A}^{n-3},
 \qquad x_A:=\sum_{a\in A}x_a.
 \label{eq:app-auxiliary-H}
\end{equation}

We take $[u]_+^0:=\Theta(u)$ off its wall and use the one-sided continuation
of Appendix~A on the wall.

We reserve $H_n:=\mathscr H_P(\beta)$ for the physical slice used in the
Letter and define the corresponding family of cut-box slices by

\begin{equation}
 \mathcal W_n(u):=
 \left\{(t_a)_{a\in P}:0\leq t_a\leq x_a,
 \ \sum_{a\in P}t_a=u\right\}.
 \label{eq:app-W-slice}
\end{equation}

We use the delta-function, equivalently coarea, normalization of the slice
volume:

\begin{equation}
 \operatorname{Vol}(\mathcal W_n(u))
 :=\int_{\prod_{a\in P}[0,x_a]}\!\dd^{n-2}t\,
 \delta\!\left(u-\sum_{a\in P}t_a\right).
 \label{eq:app-volume-normalization}
\end{equation}

Equivalently, this is the intrinsic Euclidean $(n-3)$-volume divided by the
normal length $\sqrt{n-2}$; $\operatorname{Vol}$ has this meaning throughout.

Inclusion--exclusion, or equivalently inverse Laplace transformation of
$s^{-(n-2)}\prod_{a\in P}(1-e^{-sx_a})$, gives

\begin{equation}
 \boxed{
 \mathscr H_P(u)=(n-3)!\operatorname{Vol}(\mathcal W_n(u)).}
 \label{eq:app-box-slice}
\end{equation}

In particular, writing $\mathcal W_n:=\mathcal W_n(\beta)$,

\begin{equation}
 H_n=(n-3)!\operatorname{Vol}(\mathcal W_n).
 \label{eq:app-H-volume}
\end{equation}

\subsection{Collapse of the flag sum}

Set $m=|P|=n-2$, choose $P=\{a_1,\ldots,a_m\}$, and put
$X_\ell:=x_{a_\ell}$.  In this subsection we identify $a_\ell$ with
$\ell\in[m]:=\{1,\ldots,m\}$ and write
$X_B:=\sum_{\ell\in B}X_\ell$, so $x_P=X_{[m]}$.  For
$(B_1,\ldots,B_\ell)\in\OP(U)$ with $U\subseteq[m]$, define

\begin{equation}
 s_t:=X_{B_1}+\cdots+X_{B_t},
 \qquad s_0:=0.
 \label{eq:app-positive-prefix}
\end{equation}

Maintain local momentum conservation while varying the slice position by
defining

\begin{equation}
 \Phi_m(z;X_1,\ldots,X_m):=
 \overline F\bigl(-(x_P-z),-z;P\bigr).
 \label{eq:app-Phi-definition}
\end{equation}

Specializing Eq.~\eqref{eq:app-F-flag} then gives

\begin{align}
 \Phi_m(z;X_1,\ldots,X_m)
 ={}&-|z|^{m-1}
 +\sum_{\varnothing\ne U\subseteq[m]}
 \sum_{(B_1,\ldots,B_\ell)\in\OP(U)}
 (-1)^{\ell+1}\operatorname{sgn}(-z)
 \operatorname{sgn}(s_\ell-z)|z|^{|B_1|-1}\notag\\
 &\quad
 \times
 \left(\prod_{t=2}^{\ell}|s_{t-1}-z|^{|B_t|}\right)
 |s_\ell-z|^{m-|U|}.
 \label{eq:app-Phi}
\end{align}

The two endpoint orders in Eq.~\eqref{eq:app-h-from-F} are

\begin{equation}
 \overline F(k_1,k_2;P)=\Phi_m(\beta;X_1,\ldots,X_m),
 \qquad
 \overline F(k_2,k_1;P)=\Phi_m(\alpha;X_1,\ldots,X_m).
 \label{eq:app-Phi-orders}
\end{equation}

The key identity is

\begin{equation}
 \boxed{
 \Phi_m(z;X_1,\ldots,X_m)
 =-2\prod_{\ell=1}^{m}\Delta_{X_\ell}\tp{z}^{m-1},
 \qquad
 (\Delta_xf)(z):=f(z)-f(z-x).}
 \label{eq:app-Phi-difference}
\end{equation}

Here is a short inductive proof.  At fixed $z,X_1,\ldots,X_{m-1}$, group
Eq.~\eqref{eq:app-Phi} by flags obtained by deleting label $m$.  The deleted
label lies in the terminal tail, an existing block, or a singleton block.
The preimages of one reduced flag need not cancel separately.  After summing
over all reduced flags, however, adjacent singleton placements cancel the
internal-prefix derivatives pairwise.  At fixed chamber signs only the
last-prefix derivative remains, with total multiplicity $m-1$; direct
differentiation of Eq.~\eqref{eq:app-Phi} therefore gives

\begin{equation}
 \frac{\partial}{\partial X_m}\Phi_m(z;X_1,\ldots,X_m)
 =(m-1)\Phi_{m-1}(z-X_m;X_1,\ldots,X_{m-1}),
 \qquad
 \Phi_m(z;X_1,\ldots,X_{m-1},0)=0.
 \label{eq:app-flag-derivative}
\end{equation}

The second relation denotes the $X_m\to0^+$ limit of the complete grouped
sum, not a termwise assignment of signs on the wall.

Integrating Eq.~\eqref{eq:app-flag-derivative} from $0$ to $X_m$ gives

\begin{equation}
 \Phi_m(z;X_1,\ldots,X_m)
 =(m-1)\int_0^{X_m}\!\dd t\,
 \Phi_{m-1}(z-t;X_1,\ldots,X_{m-1}).
 \label{eq:app-flag-insertion}
\end{equation}

The base case is

\begin{equation}
 \Phi_1(z;X_1)
 =-1+\operatorname{sgn}(-z)\operatorname{sgn}(X_1-z)
 =-2\Delta_{X_1}\tp{z}^{0},
 \label{eq:app-Phi-base}
\end{equation}

understood inside an open chamber.  If Eq.~\eqref{eq:app-Phi-difference}
holds at $m-1$, Eq.~\eqref{eq:app-flag-insertion} and

\begin{equation}
 (m-1)\int_0^{X_m}\!\dd t\,\tp{z-t}^{m-2}
 =\tp{z}^{m-1}-\tp{z-X_m}^{m-1}
\end{equation}

add precisely the last finite difference $\Delta_{X_m}$.  This proves
Eq.~\eqref{eq:app-Phi-difference}.  Expanding its commuting differences gives

\begin{equation}
 \Phi_m(z;X_1,\ldots,X_m)=-2\mathscr H_P(z).
 \label{eq:app-Phi-H}
\end{equation}

Equations~\eqref{eq:app-h-from-F}, \eqref{eq:app-Phi-orders}, and
\eqref{eq:app-Phi-H} imply

\begin{equation}
 h_{12}^{(n)}=\mathscr H_P(\beta)+\mathscr H_P(\alpha).
 \label{eq:app-two-slices}
\end{equation}

The reflection $t_a\mapsto x_a-t_a$ maps the slice at $u=\beta$ to the
slice at $u=x_P-\beta=\alpha$ and preserves the measure in
Eq.~\eqref{eq:app-volume-normalization}.  Therefore
$\mathscr H_P(\alpha)=\mathscr H_P(\beta)=H_n$, and

\begin{equation}
 \boxed{
 h_{12}^{(n)}=2H_n
 =2(n-3)!\operatorname{Vol}(\mathcal W_n).}
 \label{eq:app-contact-H}
\end{equation}

An overall parity transformation proves the same statement for two positive
marked legs and negative spectators.  For a mixed pair, the correct object
remains the oriented flag sum in Eq.~\eqref{eq:app-explicit-flag-volume}.

\section{Cancellation in the two-minus tree sum}
\label{sec:app-two-minus}

We now impose the on-shell conditions of Eq.~\maineqref{eq:on-shell}.  Throughout
this appendix, $N=\{1,2\}\sqcup P$, $k_1=-\alpha$, $k_2=-\beta$, and
$k_a=x_a>0$ for $a\in P$, so that $x_P=\alpha+\beta$.  For every labelled
block $B$, retain the notation of the Letter,

\begin{equation}
 k_B:=\sum_{i\in B}k_i,\qquad
 w_B:=\sum_{i\in B}w_i,\qquad
 p_B:=(k_B,w_B),\qquad
 D_B:=\frac{w_B^2}{|k_B|}-1.
 \label{eq:app-block-data}
\end{equation}

The BG manipulations are performed at generic kinematics with $k_B\ne0$ for
every internal block.  If an intermediate $D_B$ vanishes, the identities
below mean the continuous limit of the fully combined numerator after the
common factor has been cancelled; no individual singular graph is evaluated
in isolation.

\subsection{One-minus current and the central cut}

Suppose the rooted configuration formed by $B$ and the root $-p_B$ contains
exactly one negative-momentum branch.  This includes either an all-plus block
with $k_B>0$ (the root is the minus branch), or a block containing one minus
leg with $k_B<0$ (the root is positive).  The exact rooted current, including
its off-shell root propagator $D_B^{-1}$, is

\begin{equation}
 \boxed{J(B)=k_B^{|B|-1}.}
 \label{eq:app-one-minus-current}
\end{equation}

To prove this, normalize $J(\{i\})=1$ and write the phase-stripped BG
recursion as

\begin{equation}
 J(B)=\frac1{D_B}
 \sum_{r=2}^{|B|}\ \sum_{\pi\in\operatorname{Part}_r(B)}
 \cV_{r+1}\bigl(-p_B,(p_C)_{C\in\pi}\bigr)
 \prod_{C\in\pi}J(C),
 \label{eq:app-current-recursion}
\end{equation}

where $\operatorname{Part}_r(B)$ denotes unordered partitions into $r$
nonempty blocks.  Assuming Eq.~\eqref{eq:app-one-minus-current} for every
proper block, its amputated numerator is

\begin{equation}
 \mathcal N(B):=
 \sum_{r=2}^{|B|}\ \sum_{\pi\in\operatorname{Part}_r(B)}
 \cV_{r+1}\bigl(-p_B,(p_C)_{C\in\pi}\bigr)
 \prod_{C\in\pi}k_C^{|C|-1}.
 \label{eq:app-current-numerator}
\end{equation}

Expand the fixed-pair coefficients of the central vertex using
Eqs.~\eqref{eq:app-ordered-E}--\eqref{eq:app-action-kernel-bridge}.  Fix a
partition with $r$ child blocks and place $C_0$ in the $b$-marked slot of one
directed kernel.  Order the other $r-1$ blocks as
$\tau=(C_{\tau(1)},\ldots,C_{\tau(r-1)})$.  Its $q$th prefix contracts

\begin{equation}
 A_q:=C_0\cup\bigcup_{\ell=1}^{q}C_{\tau(\ell)},
 \qquad 1\leq q<r-1.
 \label{eq:app-contracted-prefix}
\end{equation}

The cancellation is between dressed prefix sectors, not individual current
monomials.  For fixed $A:=A_q\subsetneq B$, sum over $C_0$, all block
refinements of $A$, and their rooted topologies.  Summing also over all marked
and unmarked placements of the $A$-subcurrent root reconstructs its full BG
numerator.  For $|A|>1$, the induction hypothesis and
Eq.~\eqref{eq:app-current-recursion} give

\begin{equation}
 \sum_{s=2}^{|A|}\ \sum_{\rho\in\operatorname{Part}_s(A)}
 \cV_{s+1}\bigl(-p_A,(p_C)_{C\in\rho}\bigr)
 \prod_{C\in\rho}J(C)=D_AJ(A).
 \label{eq:app-dressed-prefix}
\end{equation}

The contracted edge supplies $D_A^{-1}$, leaving $J(A)$ (immediate for
$|A|=1$).  For each $(q+1)$-block refinement with $C_0$ fixed in the marked
slot, the other $q$ blocks have $q!$ spectator orders, cancelling $1/q!$.
There is no marked-slot factor: the exchanged marked order is the separate
directed term in Eq.~\eqref{eq:app-action-kernel-bridge}.  Thus the $q$th
subtraction in Eq.~\eqref{eq:app-ordered-E} cancels the leading term for
the directed kernel with $A_q$ in its $b$-slot and spectators
$C_{\tau(q+1)},\ldots,C_{\tau(r-1)}$, with the same outside currents.
Every proper prefix cancels, leaving only the marked- and unmarked-root
endpoint placements:

\begin{equation}
 \mathcal N(B)
 =\frac{w_B^2}{|k_B|}k_B^{|B|-1}-k_B^{|B|-1}
 =D_Bk_B^{|B|-1}.
 \label{eq:app-current-telescope}
\end{equation}

Cancelling the common $D_B$ in Eq.~\eqref{eq:app-current-recursion} proves
Eq.~\eqref{eq:app-one-minus-current}; the root was not put on shell.

Orient the path from leg 1 to leg 2.  At each successive vertex the running
spatial momentum increases by the strictly positive total momentum of the
plus-only branches attached there.  It therefore forms a strictly increasing
sequence from $k_1<0$ to $-k_2>0$ and, away from subset walls, has a unique
sign-changing vertex.  Cutting there gives one
$\pi\in\Adm_r(N)$; conversely, the central vertex, the partition, and one
rooted topology in each block reconstruct a unique labelled tree.  Hence no
tree or symmetry factor is missed, and reattaching the exact currents gives

\begin{equation}
 A_n^{(--+\cdots+)}
 =\sum_{r=3}^{n}\sum_{\pi\in\Adm_r(N)}
 \cV_r\bigl((p_C)_{C\in\pi}\bigr)
 \prod_{C\in\pi}k_C^{|C|-1},
 \label{eq:app-central-sum}
\end{equation}

which is Eq.~\maineqref{eq:central-tree-sum}.  Expanding the block frequencies
before using $\sum_iw_i=0$ defines

\begin{equation}
    A_n^{(--+\cdots+)}=\sum_{i<j}w_iw_jG_{ij},
\end{equation}
 
\begin{equation}
 G_{ij}^{[r]}=
 \sum_{\substack{\pi\in\Adm_r(N)\\C(i)\ne C(j)}}
 h_{C(i)C(j)}^{(r)}\bigl((k_C)_{C\in\pi}\bigr)
 \prod_{C\in\pi}k_C^{|C|-1},
 \qquad G_{ij}=\sum_{r=3}^{n}G_{ij}^{[r]}.
 \label{eq:app-G-definition}
 \end{equation}

Here $C(i)$ is the unique block of $\pi$ containing $i$.  Labels in the same
block do not contribute because the current
Eq.~\eqref{eq:app-one-minus-current} is frequency independent.

\subsection{The minus--minus coefficient}

Let $m=n-2$ and introduce one commuting variable $z_a$ for each
plus label $a\in P$, retaining coefficients multilinear in
these variables.  Write
$z_B:=\prod_{a\in B}z_a$, and let $[z_B]$ denote coefficient
extraction.  With $x_a=k_a>0$, the implicit
series

\begin{equation}
 \Xi(z)=\sum_{a\in P}z_a e^{x_a\Xi(z)}
 \label{eq:app-perturbiner}
\end{equation}

generates the rooted currents:
$[z_B]\Xi=x_B^{|B|-1}$, where $x_B=\sum_{a\in B}x_a$.  If

\begin{equation}
 \mathcal D=1-\sum_{a\in P}x_az_ae^{x_a\Xi},
\end{equation}

the multivariate Lagrange--Good formula~\cite{Good1960} also gives

\begin{equation}
 [z_B]\frac{e^{-\gamma\Xi}}{\mathcal D}
 =(x_B-\gamma)^{|B|}.
 \label{eq:app-endpoint-species}
\end{equation}

The same series generates the two minus-leg blocks.  Set

\begin{equation}
 \Xi_t:=\Xi(z_ae^{-tx_a}),\qquad
 \mathcal D_t:=\mathcal D(z_ae^{-tx_a}),\qquad
 U:=\Xi-\Xi_t.
 \label{eq:app-shifted-series}
\end{equation}

For nonempty $B\subseteq P$, Eqs.~\eqref{eq:app-perturbiner} and
\eqref{eq:app-endpoint-species} give

\begin{equation}
 [z_B]\Xi_t=e^{-tx_B}x_B^{|B|-1},\qquad
 [z_B]U=(1-e^{-tx_B})x_B^{|B|-1},\qquad
 [z_B]\frac{e^{-\gamma\Xi_t}}{\mathcal D_t}
 =e^{-tx_B}(x_B-\gamma)^{|B|}.
 \label{eq:app-current-dictionary}
\end{equation}

Reflection of the cut box allows the inverse-Laplace representation

\begin{equation}
 H_n=(m-1)!\int\frac{\dd t}{2\pi\ii}\,
 \frac{e^{t\alpha}}{t^m}
 \prod_{a\in P}(1-e^{-tx_a}),
 \label{eq:app-H-Laplace}
\end{equation}

with a Bromwich contour to the right of the origin.  If the central vertex
has exactly $s$ plus-only blocks, coefficient extraction from
Eq.~\eqref{eq:app-central-sum} gives

\begin{equation}
 G_{12}^{[s+2]}=2\mathcal R_{m,s},\qquad
 \mathcal R_{m,s}:=(s-1)!\int\frac{\dd t}{2\pi\ii}\,
 \frac{e^{t\alpha}}{t^s}[z_P]
 \frac{e^{-\alpha\Xi_t}}{\mathcal D_t}
 \frac{e^{-\beta\Xi}}{\mathcal D}\frac{U^s}{s!}.
 \label{eq:app-fixed-R}
\end{equation}

The endpoint factors and $U^s/s!$ distribute the labels among two
distinguished currents and $s$ unordered nonempty plus blocks.  The factor
$2$ is the fixed-pair coefficient and $(s-1)!/t^s$ the inverse-Laplace
kernel, so every labelled cut occurs once.  Introduce
$\mathfrak R_m(\tau):=\sum_{s=1}^{m}\tau^s\mathcal R_{m,s}$.  Since
$(s-1)!/s!=1/s$,

\begin{equation}
 \mathfrak R_m(\tau)
 =-\int\frac{\dd t}{2\pi\ii}\,e^{t\alpha}[z_P]
 \frac{e^{-\alpha\Xi_t}}{\mathcal D_t}
 \frac{e^{-\beta\Xi}}{\mathcal D}
 \log\!\left(1-\frac{\tau U}{t}\right).
 \label{eq:app-log-R}
\end{equation}

Make the source change $y_a=z_ae^{x_a\Xi}$ and put
$Y=\sum_ay_a=\Xi$.  For the square-free coefficient,

\begin{equation}
 [z_P]F(z)=[y_P]F(z(y))e^{x_PY}\mathcal D.
 \label{eq:app-source-Jacobian}
\end{equation}

The factor $\mathcal D$ cancels, and
$x_P=\alpha+\beta$ reduces the exponential to $e^{\alpha U}$.  Moreover,

\begin{equation}
 U=\sum_{a\in P}y_a\bigl(1-e^{-x_a(t+U)}\bigr),\qquad
 \mathcal D_t=1-\sum_{a\in P}x_ay_ae^{-x_a(t+U)}.
\end{equation}

Lagrange inversion now yields, for any formal series $F$,

\begin{equation}
 [y_P]\frac{F(U)}{\mathcal D_t}
 =m![\lambda^m]F(\lambda)
 \prod_{a\in P}\bigl(1-e^{-x_a(t+\lambda)}\bigr).
 \label{eq:app-labelled-Lagrange}
\end{equation}

Applying Eq.~\eqref{eq:app-labelled-Lagrange} with
$F(\lambda)=e^{\alpha\lambda}[-\log(1-\tau\lambda/t)]$ gives

\begin{equation}
 \mathfrak R_m(\tau)=m!\int\frac{\dd t}{2\pi\ii}
 [\lambda^m]e^{\alpha(t+\lambda)}
 \left[-\log\!\left(1-\frac{\tau\lambda}{t}\right)\right]
 \prod_{a\in P}(1-e^{-x_a(t+\lambda)}).
 \label{eq:app-R-after-Lagrange}
\end{equation}

With $v=t+\lambda$, every factor except the logarithm depends only on $v$,
whereas

\begin{equation}
 -\log\!\left(1-\frac{\tau\lambda}{v-\lambda}\right)
 =\log(1-\lambda/v)-\log(1-(1+\tau)\lambda/v),\qquad
 [\lambda^m]\left[
 \log(1-\lambda/v)-\log(1-(1+\tau)\lambda/v)\right]
 =\frac{(1+\tau)^m-1}{mv^m}.
 \label{eq:app-log-coefficient}
\end{equation}

This formal Laurent-series identity requires no contour deformation; the
remaining $v$ residue is Eq.~\eqref{eq:app-H-Laplace}.  Therefore

\begin{equation}
 \sum_{s=1}^{m}\tau^sG_{12}^{[s+2]}
 =2\bigl[(1+\tau)^m-1\bigr]H_n,
\end{equation}

and hence

\begin{equation}
 \boxed{
 G_{12}^{[s+2]}=2\binom{n-2}{s}H_n,\qquad
 G_{12}=(2^{n-1}-2)H_n.}
 \label{eq:app-G12}
\end{equation}

\subsection{Mixed and plus--plus frequency pairs}

For $\rho>0$, the exact positive-momentum boundary mode obeys

\begin{equation}
 \Psi_\rho[\xi](x)=e^{\ii\rho x+\rho\xi(x)},
 \qquad
 G(\xi)\Psi_\rho[\xi]=-\ii\partial_x\Psi_\rho[\xi].
 \label{eq:app-harmonic-mode}
\end{equation}

On a multilinear coefficient of total positive momentum $A$, both
$G(0)=|\partial_x|$ and $-\ii\partial_x$ act by $A$, yielding the first
identity below.  For two positive marked slots, differentiating $t$ unmarked
insertions gives $t!$; the two slot orders and Fourier convention supply
$2(-1)^t$.  Thus

\begin{equation}
 \begin{aligned}
 h_{-\rho,\kappa}^{(s+2)}
 (-\rho,\kappa;q_1,\ldots,q_s)&=0,
 &q_\ell>0,\quad \rho&=\kappa+\sum_{\ell=1}^{s}q_\ell,\\
 h_{\kappa,\lambda}^{(t+3)}
 (\kappa,\lambda;-\rho,q_1,\ldots,q_t)
 &=2(-1)^t t!\prod_{\ell=1}^{t}q_\ell,
 &q_\ell>0,\quad \rho&=\kappa+\lambda+\sum_{\ell=1}^{t}q_\ell.
 \end{aligned}
 \label{eq:app-local-harmonic-identities}
\end{equation}

Here $\rho,\kappa,\lambda>0$ are the indicated block momenta, with
$s\geq1$ and $t\geq0$.  These identities may be dressed
by the exact currents without changing their label multiplicities: introduce
square-free variables and replace every rooted component by

\begin{equation}
 \zeta_{k_B}e^{\ii k_Bx}
 \longmapsto z_BJ(B)e^{\ii k_Bx}
 =z_Bk_B^{|B|-1}e^{\ii k_Bx}.
 \label{eq:app-current-replacement}
\end{equation}

Square-free extraction forbids repeated labels, and the factorial in
Eq.~\eqref{eq:app-local-harmonic-identities} cancels the exponential
coefficient; every unordered labelled partition therefore has unit weight.

Fix the negative block $C_1\ni1$, write $k_{C_1}=-\rho$, and keep a plus
label $a\notin C_1$ marked.  In the square-free coefficient of the exact
identity

\begin{equation}
 \left\langle e^{-\ii Ax},
 [G(\xi)-G(0)]\Psi_\rho[\xi]\right\rangle=0
\end{equation}

$a$ is supplied either by a nonlinear slot of $G(\xi)$ at the central
vertex or by the exponential in the current on the leg-2 side.  Every
admissible partition with fixed $C_1$ occurs once in exactly one class, so
the identity pairs the two complete class sums, not individual graphs.
Summing over $C_1$ and then exchanging legs 1 and 2 gives

\begin{equation}
 \boxed{G_{1a}=G_{2a}=0.}
 \label{eq:app-mixed-zero}
\end{equation}

For two plus labels $a,b$, self-adjointness in the spatial bilinear pairing
gives the unrestricted identity

\begin{equation}
 2\langle\Psi_{k_a},G(\xi)\Psi_{k_b}\rangle
 =-\ii\int\dd x\,\partial_x
 (\Psi_{k_a}\Psi_{k_b})=0.
 \label{eq:app-unrestricted-plus}
\end{equation}

For periodic or spatially decaying fields the boundary term vanishes; after
the current replacement call this coefficient $\mathcal U_{ab}=0$.  Let
$\mathcal S_{12}$ contain partitions with both minus labels in one block.
Among partitions in which legs 1 and 2 lie in distinct blocks, let
$\mathcal C_{\alpha|\beta}$ or $\mathcal C_{\beta|\alpha}$ contain those with
only the leg-1 or leg-2 block negative.  These classes are disjoint.  No class
with neither block negative exists because all other blocks are positive and
total block momentum vanishes. 

Hence,
\begin{equation}
 G_{ab}=\mathcal U_{ab}-\mathcal S_{12}
 -\mathcal C_{\alpha|\beta}-\mathcal C_{\beta|\alpha}.
 \label{eq:app-physical-plus-decomposition}
\end{equation}

Put $x:=k_a$, $y:=k_b$,
$U:=P\setminus\{a,b\}$, and $d:=|U|+1=n-3$.  Write
$(\nabla_U f)(c):=\sum_{T\subseteq U}(-1)^{|T|}f(c-k_T)$ and
$\mathscr B_c:=\tp{c-x}^{d}+\tp{c-y}^{d}-\tp{c-x-y}^{d}$.

In inverse-Laplace form an unmarked positive component contributes
$1-e^{-\eta k_u}$ according as it is exposed or absorbed, while the marked
exposed sets $\{a\},\{b\},\{a,b\}$ contribute
$e^{-\eta x}+e^{-\eta y}-e^{-\eta(x+y)}$.  Define

\begin{equation}
 \Phi_U(\eta):=\prod_{u\in U}(1-e^{-\eta k_u}),
 \qquad
 M_{xy}(\eta):=e^{-\eta x}+e^{-\eta y}-e^{-\eta(x+y)}.
 \label{eq:app-plus-Laplace-data}
\end{equation}

Square-free coefficient extraction gives
\begin{align}
 \nabla_U\tp{c}^{d}
 =d!\int_{\mathrm{Br}}\frac{\dd\eta}{2\pi\ii}\,
 \frac{e^{\eta c}}{\eta^{d+1}}\Phi_U(\eta),
 \nabla_U\mathscr B_c
 =d!\int_{\mathrm{Br}}\frac{\dd\eta}{2\pi\ii}\,
 \frac{e^{\eta c}}{\eta^{d+1}}\Phi_U(\eta)M_{xy}(\eta),
 \label{eq:app-plus-Laplace-inversion}
\end{align}
where $\mathrm{Br}$ is a Bromwich contour.  The two marked-slot orders supply
the overall factor 2, so the one-negative-block classes are
\begin{align}
 \mathcal C_{\alpha|\beta}
 =2d!\int_{\mathrm{Br}}\frac{\dd\eta}{2\pi\ii}\,
 \frac{\Phi_U(\eta)}{\eta^{d+1}}e^{\eta\alpha}M_{xy}(\eta),
 \mathcal C_{\beta|\alpha}
 =2d!\int_{\mathrm{Br}}\frac{\dd\eta}{2\pi\ii}\,
 \frac{\Phi_U(\eta)}{\eta^{d+1}}e^{\eta\beta}M_{xy}(\eta).
 \label{eq:app-one-negative-Laplace}
\end{align}

For $\mathcal U_{ab}-\mathcal S_{12}$, set $Q:=\{1,2\}\cup U$,
$(\kappa_1,\kappa_2)=(-\alpha,-\beta)$, $\kappa_u=k_u$, and use $y$ as the
marked reference.  A subset $T\subseteq Q$ records components absorbed into
its boundary mode.  Equation~\eqref{eq:app-local-harmonic-identities} and
inclusion--exclusion give the first line below; the second subtracts the
merged-minus sector:
\begin{align}
 \mathcal U_{ab}-\mathcal S_{12}
 ={}&(-1)^d\sum_{T\subseteq Q}(-1)^{|T|}
 |y+\kappa_T|(y+\kappa_T)^{d-1}+2\sum_{T\subseteq U}(-1)^{|T|}
 (\alpha+\beta-k_T)^d .
 \label{eq:app-unrestricted-minus-same-block}
\end{align}
The second-line argument is positive because
$\alpha+\beta-k_T=x+y+k_{U\setminus T}>0$. Using
\begin{equation}
 |z|z^{d-1}=\tp{z}^{d}+(-1)^{d-1}\tp{-z}^{d},
 \qquad x+y+k_U=\alpha+\beta,
 \label{eq:app-positive-negative-split}
\end{equation}
and complementing $T\mapsto U\setminus T$ in the negative-part terms gives
\begin{align}
 \mathcal U_{ab}-\mathcal S_{12}
 =2\nabla_U\left(\mathscr B_\alpha+\tp{\beta}^{d}\right)=2d!\int_{\mathrm{Br}}\frac{\dd\eta}{2\pi\ii}\,
 \frac{\Phi_U(\eta)}{\eta^{d+1}}
 \left[e^{\eta\alpha}M_{xy}(\eta)+e^{\eta\beta}\right].
 \label{eq:app-unrestricted-boundary-Laplace}
\end{align}
Thus $e^{\eta\beta}$ is the Laplace image of the uncancelled endpoint
$\nabla_U\tp{\beta}^{d}$, not an extra configuration.  After removing
$2d!\Phi_U(\eta)/\eta^{d+1}$, Eq.~\eqref{eq:app-physical-plus-decomposition}
reduces to
$[e^{\eta\alpha}M_{xy}+e^{\eta\beta}]-e^{\eta\alpha}M_{xy}
-e^{\eta\beta}M_{xy}=e^{\eta\beta}(1-e^{-\eta x})(1-e^{-\eta y})$.

Multiplying back the common factor and inverse transforming gives

\begin{align}
 G_{ab}
 &=2\nabla_U\left(
 \tp{\beta}^{d}-\tp{\beta-x}^{d}-\tp{\beta-y}^{d}
 +\tp{\beta-x-y}^{d}\right)\notag\\
 &=2\sum_{A\subseteq P}(-1)^{|A|}\tp{\beta-k_A}^{n-3}
 =2H_n.
 \label{eq:app-plus-plus}
\end{align}

All propagator denominators and apparent inverse subset-momentum factors have
cancelled before the limit.  The identities extend to chamber walls by
one-sided continuation of the complete piecewise-polynomial expressions.

Together, Eqs.~\eqref{eq:app-G12}, \eqref{eq:app-mixed-zero}, and
\eqref{eq:app-plus-plus} prove Eq.~\maineqref{eq:three-identities}.  Frequency
conservation then gives the amplitude in Eq.~\maineqref{eq:amplitude}.

\end{document}